\documentclass{article}
\usepackage{amssymb}
\usepackage{amsmath}
\usepackage{geometry}
\usepackage{units}
\usepackage{graphicx}
\usepackage{amsfonts}
\usepackage{float}
\usepackage{bm}
\usepackage{color}
\usepackage{epstopdf}
\usepackage{subcaption}
\usepackage{cite}

\begin{document}

\title{Effects of Extended Uncertainty Principle on the Relativistic Coulomb
Potential}
\author{B. Hamil \thanks{hamilbilel@gmail.com} \\
D\'{e}partement de TC de SNV, Universit\'{e} Hassiba Benbouali, Chlef,
Algeria. \  \and M. Merad \thanks{meradm@gmail.com} \\
Facult\'{e} des Sciences Exactes, Universit\'{e} de Oum El Bouaghi, 04000
Oum El Bouaghi, Algeria. \and T. Birkandan \thanks{birkandant@itu.edu.tr} \\
Department of Physics, Istanbul Technical University, 34469 Istanbul,
Turkey. }
\date{}
\maketitle

\begin{abstract}
The relativistic bound-state energy spectrum and the wavefunctions for the
Coulomb potential are studied for de Sitter and anti-de Sitter spaces in the
context of the extended uncertainty principle. Klein-Gordon and Dirac
equations are solved analytically to obtain the results. The electron
energies of hydrogen-like atoms are studied numerically.
\end{abstract}

\begin{description}
\item[PACS numbers:] 02.40.Gh, 03.65.Ge, 03.65.Pm.

\item[Keywords:] Extended uncertainty principle, Modified commutation
relations, minimal uncertainty in momentum, (A)dS space, Relativistic
Coulomb Potential.
\end{description}

\section{Introduction}

The standard Heisenberg uncertainty principle (HUP) \cite{1} of quantum
mechanics represents one of the fundamental properties of quantum systems.
According to HUP, there should be a fundamental limit for the measurement
accuracy where certain pairs of physical observables, such as the positions
and momenta or energy and time, cannot be simultaneously measured with full
accuracy. A large number of studies have converged on the idea that the HUP
should be reformulated for systems with energies close to the Planck scale $%
\kappa $, which incorporates the concept of minimum measurable length \cite%
{2,3,4,5,6}. The minimum measurable length idea is predicted by different
tentative approaches to quantum gravity such as string theory \cite{7,8},
quantum geometry \cite{9,10}, loop quantum gravity \cite{11} and black hole
physics \cite{12,13}. This fundamental scale leads to a modification of the
Heisenberg uncertainty principle to the so-called generalized uncertainty
principle (GUP) \cite{14,15,16}. The GUP ideas are characterized by a
deformation of the classical Heisenberg uncertainty relation and the most
widely adopted generalization of the Heisenberg uncertainty principle reads 
\cite{14}, 
\begin{equation}
\left( \Delta X\right) \left( \Delta P\right) \geqslant \frac{\hbar }{2}%
\left( 1+l_{p}\left( \Delta P\right) ^{2}\right) .
\end{equation}
If $l_{p}$ is a positive constant, the formula implies the existence of a
minimal momentum uncertainty. If $l_{p}$ is a negative constant, no minimal
uncertainty occurs.

On the other hand, there exists another form of the deformed Heisenberg
uncertainty principle which is called the extended uncertainty principle
(EUP) \cite{17,18,19,20,21,22} which takes into account the large distances
where, due to gravity, the spacetime is curved. We have, 
\begin{equation}
\left( \Delta X_{i}\right) \left( \Delta P_{i}\right) \geqslant \frac{\hbar 
}{2}\left( 1+\frac{\left( \Delta X_{i}\right) ^{2}}{l_{H}^{2}}\right) ,
\end{equation}%
where $l_{H}$ is the (anti-)de Sitter radius which is related to the
cosmological constant ($\Lambda $) as $\Lambda =-3/l_{H}^{2}$. The
EUP modifies the standard commutation relations between position and, the
momentum and the coordinate representation of the momentum operators for
this model become position-dependent.

S. Mignemi showed that EUP can be derived from the geometric properties of
the (anti-)de Sitter spacetime, with a suitable parametrization \cite{17}.
In addition to this, it is shown that the EUP arises naturally from the
first terms in the expansion of any metric, which means that the corrections
to the Hawking temperature of Schwarzschild black hole can be computed by
incorporating the gravitational interaction as an external force on a flat
background, and neglecting the curvature of spacetime \cite{18}. To our
knowledge, only a few works have studied the influence of extended
uncertainty principle on quantum mechanical problems \cite%
{23,24,25,26,27,28,29,30,31,32,33,34,35}.

In this paper, we study the problem of relativistic Coulomb potential in the
framework of the extended uncertainty principle in (3+1) dimensional
spacetimes. We solve the problem of Coulomb potential for the Klein-Gordon
and Dirac equations to get the exact form of the energy levels and
eigenfunctions. This paper is organized as follows: In Sect. 2, we introduce
the main relations of quantum mechanics with the extended uncertainty
principle. In sections 3 and 4, we solve the Klein-Gordon and Dirac
equations exactly in (3+1) dimensions with the Coulomb-like interaction in
the context of EUP in the position space representation. Section 5 contains
the conclusion.

\section{The Extended Uncertainty Relation}

In three-dimensional space, the modified Heisenberg algebra leading to EUP
is given by the following deformed commutation relations \cite{17}, 
\begin{eqnarray}
\left[ X_{i},P_{j}\right]  &=&i\hslash \left( \delta _{ij}-s\lambda
X_{i}X_{j}\right) ;\text{ }i=j=1,2,3,  \notag \\
\left[ X_{i},X_{j}\right]  &=&0;\text{ \  \ }\left[ P_{i},P_{j}\right]
=-i\hslash s\lambda L_{ij},  \label{3}
\end{eqnarray}%
where $\lambda =-\frac{1}{l_{H}^{2}s}$ is a small parameter of dimension of
inverse distance squared, $L_{ij}=X_{i}P_{j}-X_{j}P_{i}$, $s=1$ for de
Sitter space and $s=-1$ for anti-de Sitter space. These commutation
relations lead to the extended uncertainty principle, 
\begin{equation}
\left( \Delta X_{i}\right) \left( \Delta P_{i}\right) \geq \frac{\hslash }{2}%
\left( 1-s\lambda \left( \Delta X_{i}\right) ^{2}\right) .  \label{4}
\end{equation}%
In anti-de Sitter space $\left( s=-1\right) ,$ the uncertainty relation (\ref%
{4}) is characterized by the appearance of a non-zero minimal uncertainty in
momentum (MUM), 
\begin{equation}
\left( \Delta P_{k}\right) _{\min }=\frac{\hslash \sqrt{\lambda }}{2},\text{
\  \  \ }\forall k,
\end{equation}%
and for the case of de Sitter space $\left( s=1\right) $, no lower bound on
the measurable length arises. An explicit representation of the momentum and
position operators obeying Eq. (\ref{3}) is given by, 
\begin{eqnarray}
X_{i} &=&\frac{x_{i}}{\sqrt{1+s\lambda r^{2}}},\text{ \  \  \ where }%
r=\sum_{i=1}^{3}x_{i}^{2}\text{.} \\
P_{i} &=&\frac{\hslash }{i}\sqrt{1+s\lambda r^{2}}\frac{\partial }{\partial
x_{i}},
\end{eqnarray}%
in the position representation. In (anti-)de sitter space, the scalar
product is not the usual one, but it is defined as, 
\begin{equation}
\left \langle \phi \right. \left \vert \psi \right \rangle =\int \frac{d\vec{r}%
}{\sqrt{1+s\lambda r^{2}}}\phi ^{\dagger }\left( \vec{r}\right) \psi \left( 
\vec{r}\right) ,
\end{equation}%
which preserves the hermiticity of the position operator.

\section{Klein-Gordon Equation for the Hydrogen Atom}

In this section, we will study the eigenvalue problem of the Klein-Gordon
equation for a Coulomb-type interaction in (3+1) dimensional spacetime. We
have, 
\begin{equation}
\left[ \left( E+\frac{Ze^{2}}{R}\right) ^{2}-c^{2}P^{2}-m^{2}c^{4}\right]
\psi =0.  \label{9}
\end{equation}
In de Sitter space, the momentum squared and distance operators act in
coordinate space as%
\begin{equation}
P^{2}\psi =-\hbar ^{2}\left[ \left( 1+\lambda r^{2}\right) \left( \frac{%
\partial ^{2}}{\partial r^{2}}+\frac{2}{r}\frac{\partial }{\partial r}-\frac{%
\hat{L}^{2}}{\hslash ^{2}r^{2}}\right) +\lambda r\frac{\partial }{\partial r}%
\right] \psi ,  \label{10}
\end{equation}%
\begin{equation}
R\psi =\frac{r}{\sqrt{1+\lambda r^{2}}}\psi ,  \label{11}
\end{equation}
where $\hat{L}$ is the orbital angular momentum operator whose
eigenfunctions are given in terms of the spherical harmonics,%
\begin{equation}
\hat{L}^{2}Y_{\ell ,\nu }\left( \theta ,\varphi \right) =\hslash ^{2}\ell
\left( \ell +1\right) Y_{\ell ,\nu }\left( \theta ,\varphi \right) ,
\end{equation}
where $\ell$ and $\nu$ are quantum numbers. Using (\ref{10}) and (\ref{11})
in Eq. (\ref{9}), we obtain,%
\begin{equation}
\left[ \left( 1+\lambda r^{2}\right) \left( \frac{\partial ^{2}}{\partial
r^{2}}+\frac{2}{r}\frac{\partial }{\partial r}-\frac{\hat{L}^{2}}{\hslash
^{2}r^{2}}\right) +\lambda r\frac{\partial }{\partial r}+\frac{2Z\mu E}{%
\hbar c}\frac{\sqrt{1+\lambda r^{2}}}{r}+\left( Z\mu \right) ^{2}\frac{%
\left( 1+\lambda r^{2}\right) }{r^{2}}+\frac{E^{2}-m^{2}c^{4}}{\hbar
^{2}c^{2}}\right] \psi =0,
\end{equation}%
where $\mu =\frac{e^{2}}{\hbar c}\simeq 1/137.03602$ is Sommerfeld's
fine-structure constant and $\frac{2\pi \hbar }{m}=2.4263\times 10^{-12}$m
is the Compton wavelength. For the wave function $\psi$, we make the ansatz, 
\begin{equation}
\psi =\frac{\digamma _{\ell }\left( r\right) }{\sqrt{r}}Y_{\ell ,\nu }\left(
\theta ,\varphi \right).
\end{equation}
The angular and radial dependent parts can now be separated to yield the
radial equation,%
\begin{equation}
\left[ \left( 1+\lambda r^{2}\right) \left( \frac{d^{2}}{dr^{2}}+\frac{1}{r}%
\frac{d}{dr}-\frac{\delta ^{2}}{r^{2}}\right) +\lambda r\frac{d}{dr}+\frac{%
2Z\mu E}{\hbar c}\frac{\sqrt{1+\lambda r^{2}}}{r}+\frac{E^{2}-m^{2}c^{4}}{%
\hbar ^{2}c^{2}}-\frac{\lambda }{2}\right] \digamma _{\ell }\left( r\right)
=0,  \label{15}
\end{equation}%
with $\delta ^{2}=\left( \ell +\frac{1}{2}\right) ^{2}-\left( Z\mu \right)
^{2}.$ Let us now transform Eq. (\ref{15}) into a hypergeometric
differential equation by using two successive changes of variables as 
\begin{equation}
\varkappa =\frac{\sqrt{1+\lambda r^{2}}}{\sqrt{\lambda }r}\text{ and }y=%
\frac{1}{2}\left( 1-\varkappa \right) .
\end{equation}%
Equation (\ref{15}) takes the form, 
\begin{equation}
\left[ y\left( 1-y\right) \frac{\partial ^{2}}{\partial y^{2}}+\left( \frac{1%
}{2}-y\right) \frac{\partial }{\partial \varkappa }-\frac{\delta ^{2}-\frac{%
2Z\mu E}{\hbar c\sqrt{\lambda }}-\frac{E^{2}-m^{2}c^{4}}{\hbar
^{2}c^{2}\lambda }-\frac{1}{2}}{4y}-\frac{\delta ^{2}+\frac{2Z\mu E}{\hbar c%
\sqrt{\lambda }}-\frac{E^{2}-m^{2}c^{4}}{\hbar ^{2}c^{2}\lambda }-\frac{1}{2}%
}{4\left( 1-y\right) }+\delta ^{2}\right] \digamma _{\ell }\left( y\right)
=0.  \label{17}
\end{equation}%
The latter equation possesses three regular singular points located at $%
y=\{0,1,\infty \}$. Applying the definition,%
\begin{equation}
\digamma _{\ell }\left( y\right) =y^{a}\left( 1-y\right) ^{b}\Xi _{\ell
}\left( y\right),
\end{equation}%
the Eq. (\ref{17}) will reduce to the hypergeometric type, namely, 
\begin{equation}
\left \{ y\left( 1-y\right) \frac{\partial ^{2}}{\partial y^{2}}+\left( 
\frac{1}{2}+2a-y\left( 1+2a+2b\right) \right) \frac{\partial }{\partial y}-%
\left[ \left( a+b\right) ^{2}-\delta ^{2}\right] \right \} \Xi _{\ell
}\left( y\right) =0,  \label{19}
\end{equation}
where $a$ and $b$ are given by%
\begin{eqnarray}
a &=&\frac{1}{4}+\frac{1}{2}\sqrt{\delta ^{2}-\frac{2Z\mu E}{\hbar c\sqrt{%
\lambda }}-\frac{E^{2}-m^{2}c^{4}}{\hbar ^{2}c^{2}\lambda }+\frac{3}{4}}, \\
b &=&\frac{1}{4}+\frac{1}{2}\sqrt{\delta ^{2}+\frac{2Z\mu E}{\hbar c\sqrt{%
\lambda }}-\frac{E^{2}-m^{2}c^{4}}{\hbar ^{2}c^{2}\lambda }+\frac{3}{4}}.
\end{eqnarray}%
The regular solution at the origin $y=0$ of Eq.(\ref{19}) is written in
terms of the hypergeometric function as,%
\begin{equation}
\Xi _{\ell }\left( y\right) =\mathbf{F}\left( A;B;\frac{1}{2}+2a;y\right) ,
\end{equation}
whose parameters are given by%
\begin{equation}
A=a+b-\delta ;\text{ \ }B=a+b+\delta .
\end{equation}
The hypergeometric function becomes a polynomial of degree $n$ when%
\begin{equation}
A=-n\text{ \  \ or \ }B=-n\text{ \  \ where }n=0,1,2,...
\end{equation}%
In both cases, we have 
\begin{equation}
\frac{1}{2}+\frac{1}{2}\sqrt{\delta ^{2}+\frac{3}{4}+\frac{m^{2}c^{4}}{\hbar
^{2}c^{2}\lambda }-\frac{2Z\mu E}{\hbar c\sqrt{\lambda }}-\frac{E^{2}}{\hbar
^{2}c^{2}\lambda }}+\frac{1}{2}\sqrt{\delta ^{2}+\frac{3}{4}+\frac{m^{2}c^{4}%
}{\hbar ^{2}c^{2}\lambda }+\frac{2Z\mu E}{\hbar c\sqrt{\alpha }}-\frac{E^{2}%
}{\hbar ^{2}c^{2}\lambda }}+\delta =-n.  \label{25}
\end{equation}
The energy spectrum can be obtained from (\ref{25}) which leads to%
\begin{equation}
E_{KG}^{dS}=\frac{mc^{2}\sqrt{1-\frac{\hbar ^{2}\lambda }{m^{2}c^{2}}\left[
\left( N-\ell -\frac{1}{2}+\sqrt{\left( \ell +\frac{1}{2}\right) ^{2}-\left(
Z\mu \right) ^{2}}\right) ^{2}+\left( Z\mu \right) ^{2}-\ell \left( \ell
+1\right) -1\right] }}{\sqrt{1+\frac{Z^{2}\mu ^{2}}{\left( N-\ell -\frac{1}{2%
}+\sqrt{\left( \ell +\frac{1}{2}\right) ^{2}-\left( Z\mu \right) ^{2}}%
\right) ^{2}}}},  \label{26}
\end{equation}
where $N=n+\ell +1$ is the principal quantum number. We observe that the
energy levels depend on the deformation parameter $\lambda $. This is a
natural consequence of the modified Heisenberg algebra. We also note that
the energy levels depend on $N^{2}$, which is the feature of hard
confinement. According to the energy levels, we remark the following:

\begin{itemize}
\item The constraint $\ell +\frac{1}{2}\geqslant Z\mu $ is necessary for the
existence of physical energy eigenvalues.

\item For $\ell =0,$ we must impose $Z\leqslant 69$ and bound states do not
exist for larger $Z$-values.

\item For larger $N$-values, the energy spectrum would have an unphysical
behavior.
\end{itemize}

The expansion of Eq.(\ref{26}) up to the first order in $\lambda$ yields%
\begin{equation}
E_{KG}^{dS}=\varepsilon _{KG}-\frac{\hbar ^{2}\lambda \varepsilon _{KG}}{%
2m^{2}c^{2}}\left[ \left( N-\ell -\frac{1}{2}+\sqrt{\left( \ell +\frac{1}{2}%
\right) ^{2}-\left( Z\mu \right) ^{2}}\right) ^{2}+\left( Z\mu \right)
^{2}-\ell \left( \ell +1\right) -1\right] ,  \label{27}
\end{equation}
where, 
\begin{equation}
\varepsilon _{KG}=mc^{2}\left[ 1+\frac{Z^{2}\mu ^{2}}{\left( N-\ell -\frac{1%
}{2}+\sqrt{\left( \ell +\frac{1}{2}\right) ^{2}-\left( Z\mu \right) ^{2}}%
\right) ^{2}}\right] ^{-\frac{1}{2}}.
\end{equation}
The first term in (\ref{27}) is the energy spectrum of the ordinary
three-dimensional Coulomb potential of spin-0 particles with no deformation,
while the second term represents the correction due to the presence of the
EUP. Expanding (\ref{27}) in powers of $\left( Z\mu \right) $ yields 
\begin{eqnarray}
W_{KG}^{ds} &=&E_{KG}^{dS}-mc^{2}=-\frac{mZ^{2}e^{4}}{2\hbar ^{2}N^{2}}-%
\frac{\hbar ^{2}\lambda }{2m}\left[ N^{2}-\ell \left( \ell +1\right) -1%
\right] -mc^{2}\frac{Z^{4}\mu ^{4}}{2N^{4}}\left( \frac{N}{\ell +\frac{1}{2}}%
-\frac{3}{4}\right) \left \{ 1-\frac{\hbar ^{2}\lambda }{2m^{2}c^{2}}\left[
N^{2}-\ell \left( \ell +1\right) -1\right] \right \}  \notag \\
&&+\frac{\hbar ^{2}\lambda }{2m}\frac{\left( Z\mu \right) ^{2}}{2N^{2}}\left
\{ \left[ \frac{2N^{3}}{\left( \ell +\frac{1}{2}\right) }-N^{2}-\ell \left(
\ell +1\right) -1\right] +\left( Z\mu \right) ^{2}\left[ 1+\frac{N^{3}}{%
2\left( \ell +\frac{1}{2}\right) ^{3}}-\frac{N^{2}}{2\left( \ell +\frac{1}{2}%
\right) ^{2}}-\frac{N}{\left( \ell +\frac{1}{2}\right) }\right] \right \} .
\label{29}
\end{eqnarray}%
The first and the second terms in (\ref{29}) represent the non-relativistic
energy levels of hydrogen in dS space \cite{35}, while the other terms are
relativistic corrections in dS space.

The same calculation can be performed for the case of anti-de Sitter space
by taking the change of the sign in the deformation into the account by
using $\lambda>0$. The energy spectrum of the system is found as 
\begin{equation}
E_{KG}^{AdS}=\frac{mc^{2}\sqrt{1+\frac{\hbar ^{2}\lambda }{m^{2}c^{2}}\left[
\left( N-\ell -\frac{1}{2}+\sqrt{\left( \ell +\frac{1}{2}\right) ^{2}-\left(
Z\mu \right) ^{2}}\right) ^{2}+\left( Z\mu \right) ^{2}-\ell \left( \ell
+1\right) -1\right] }}{\sqrt{1+\frac{Z^{2}\mu ^{2}}{\left( N-\ell -\frac{1}{2%
}+\sqrt{\left( \ell +\frac{1}{2}\right) ^{2}-\left( Z\mu \right) ^{2}}%
\right) ^{2}}}}.
\end{equation}
We can plot our results for different scenarios. The contribution of the
deformation is very small numerically for physical values of the parameters.
Thus we take $\hbar=c=m=1$ in order to avoid numerical errors and make the
effect of the deformation factor visible in the graphics. In Fig. (\ref%
{fig:fig1}) we plot the ratio $\frac{E_{KG}^{(A)dS}}{\varepsilon_{KG}}$ with
respect to $N$ for some values of the deformation parameter $\lambda$. Here, 
$\ell=0$ and $Z=50$. We see that the curves for the AdS case grow with $N$,
while they decrease almost rapidly with $N$ in the dS case. The undeformed
calculation shown with a solid line with $\lambda=0$ gives the ratio as
unity as expected.

The Figures (\ref{fig:fig2}) and (\ref{fig:fig3}) show the behavior of the
energy eigenvalues $E_{KG}^{(A)dS}$ with respect to $Z$. We take $N=1$ and $%
\ell=0$ in Fig. (\ref{fig:fig2}). As we have calculated above, no bound
states exist for $Z>69$. In Fig. (\ref{fig:fig3}), for $N=3$ and $\ell=1$,
we verify the condition $\ell +\frac{1}{2}\geqslant Z\mu$ graphically by
having $Z=206$ as the accumulation point. 
\begin{figure}[tbph]
\centering
\includegraphics[scale=0.7]{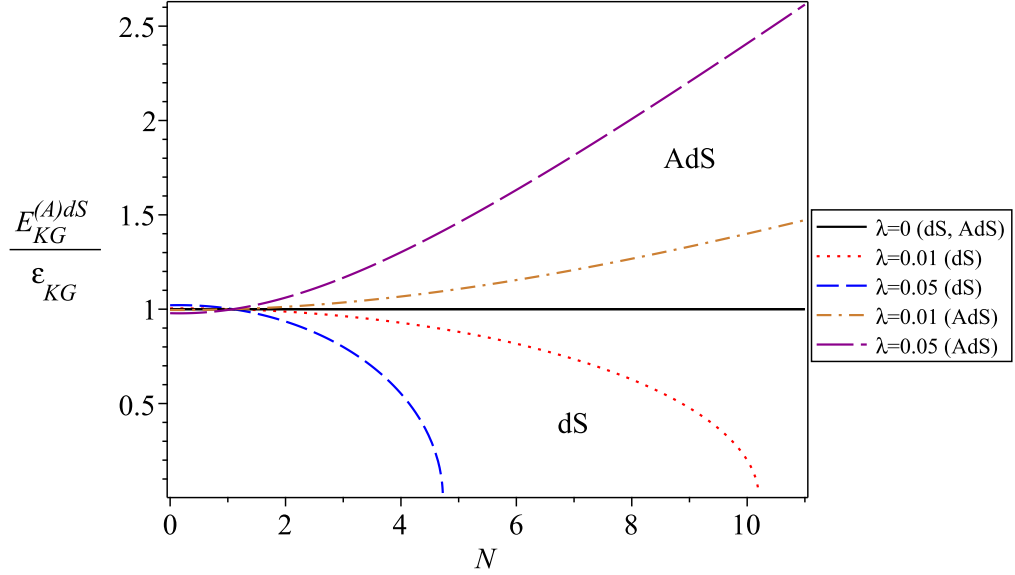}
\caption{$\frac{E_{KG}^{(A)dS}}{\protect \varepsilon_{KG}}$ vs. $N$}
\label{fig:fig1}
\end{figure}
\begin{figure}[tbph]
\centering
\includegraphics[scale=0.7]{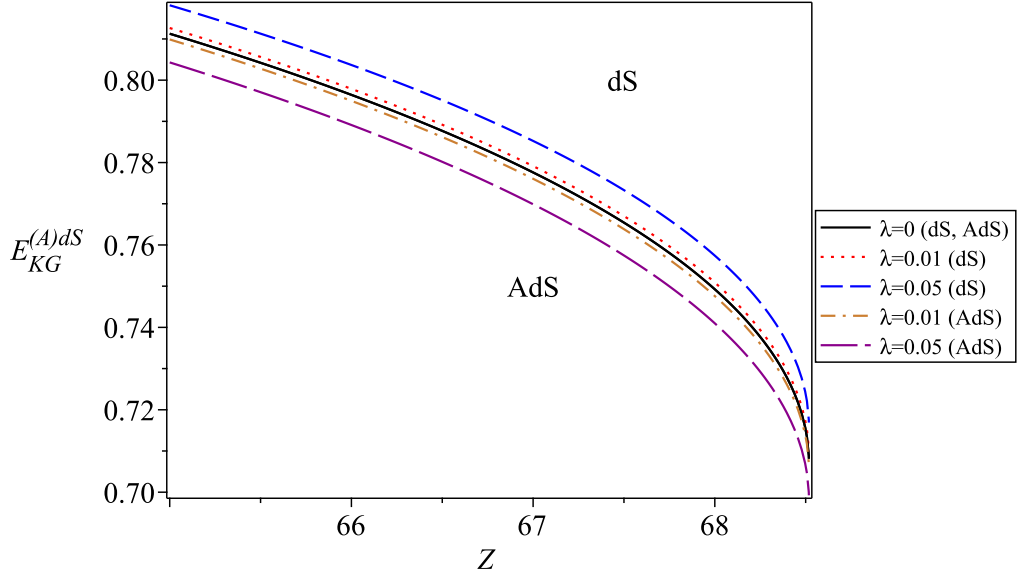}
\caption{$E_{KG}^{(A)dS}$ vs. $Z$ for $N=1$, $\ell=0$}
\label{fig:fig2}
\end{figure}
\begin{figure}[tbph]
\centering
\includegraphics[scale=0.7]{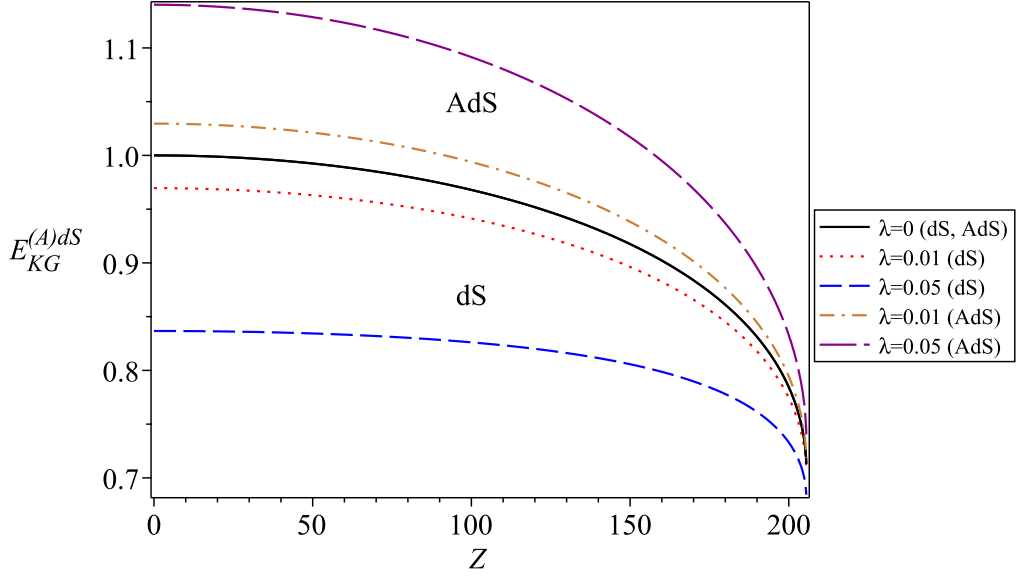}
\caption{$E_{KG}^{(A)dS}$ vs. $Z$ for $N=3$, $\ell=1$}
\label{fig:fig3}
\end{figure}

\section{Dirac Equation for the Hydrogen Atom}

The Dirac equation in the presence of the Coulomb potential $V$ is%
\begin{equation}
\left[ c\overrightarrow{\mathbf{\alpha }}\overrightarrow{\cdot P}+\beta
mc^{2}-E+V\right] \psi =0,  \label{31}
\end{equation}
where the matrices $\mathbf{\alpha }$ and $\beta $ satisfy the
anticommutation relations,%
\begin{equation}
\left \{ \mathbf{\alpha }_{i},\beta \right \} =0;\text{ \  \ }\left \{ 
\mathbf{\alpha }_{i},\mathbf{\alpha }_{j}\right \} =2\delta _{ij};\text{ \  \ 
}i,j=1,2,3.
\end{equation}
An explicit familiar representation of\ $\mathbf{\alpha }$\ and $\beta $\ is
provided by%
\begin{equation}
\mathbf{\alpha }_{i}=\left( 
\begin{array}{cc}
0 & \sigma _{i} \\ 
\sigma _{i} & 0%
\end{array}%
\right) ;\text{ \  \ }\beta =\left( 
\begin{array}{cc}
1 & 0 \\ 
0 & -1%
\end{array}%
\right) ,
\end{equation}
where $\sigma _{i}$ are $2 \times 2$ Pauli spin matrices. In spherical
coordinates, the operator $\overrightarrow{\mathbf{\alpha }}\overrightarrow{%
\cdot P}$ can be written as%
\begin{equation}
\overrightarrow{\mathbf{\alpha }}\overrightarrow{\cdot P}=\sqrt{1+\lambda
r^{2}}\overrightarrow{\mathbf{\alpha }}\overrightarrow{\cdot p}=-i\hbar 
\sqrt{1+\lambda r^{2}}\mathbf{\alpha }_{r}\left( \frac{\partial }{\partial r}%
+\frac{1}{r}-\frac{\beta }{\hbar r}\boldsymbol{K}\right) ,
\end{equation}
in the deformed case. Here, 
\begin{equation}
\mathbf{\alpha }_{r}=\overrightarrow{\mathbf{\alpha }}\overrightarrow{\cdot r%
};\text{ \  \  \ }\boldsymbol{K}=\beta \left( \overrightarrow{\sum }\cdot 
\overrightarrow{\mathbf{L}}+\hbar \right) ;\text{ \ }\overrightarrow{\sum }%
=\left( 
\begin{array}{cc}
\overrightarrow{\sigma } & 0 \\ 
0 & -\overrightarrow{\sigma }%
\end{array}%
\right) ,
\end{equation}%
where the last operator is absent in the Klein-Gordon equation. The solution
ansatz of (\ref{31}) can be chosen such that 
\begin{equation}
\psi =\left( 
\begin{array}{c}
f_{1}\left( r\right) \mathcal{Y}_{\kappa ,\nu }\left( \theta ,\varphi \right)
\\ 
if_{2}\left( r\right) \mathcal{Y}_{-\kappa ,\nu }\left( \theta ,\varphi
\right)%
\end{array}%
\right) ,
\end{equation}
where $\mathcal{Y}_{\kappa ,\nu }\left( \theta ,\varphi \right) $ are spinor
spherical harmonics. Considering the action of the operator $\boldsymbol{K}$%
, we have 
\begin{eqnarray}
\boldsymbol{K}\mathcal{Y}_{\kappa ,\nu }\left( \theta ,\varphi \right)
&=&-\hbar \kappa \mathcal{Y}_{\kappa ,\nu }\left( \theta ,\varphi \right) , 
\notag \\
\boldsymbol{K}\mathcal{Y}_{-\kappa ,\nu }\left( \theta ,\varphi \right)
&=&\hbar \kappa \mathcal{Y}_{-\kappa ,\nu }\left( \theta ,\varphi \right) ,
\end{eqnarray}
where the quantum number $\kappa$ is defined as%
\begin{equation}
\kappa =\pm \left( j+\frac{1}{2}\right) =\left \{ 
\begin{array}{c}
-\left( \ell +1\right) \text{ \  \  \  \ for }j=l+1/2 \\ 
\ell \text{ \  \  \  \  \  \  \  \  \  \  \  \  \  \  \ for }j=l-1/2%
\end{array}%
\right. .
\end{equation}
Using the expression for $\overrightarrow{\mathbf{\alpha }}\overrightarrow{%
\cdot P}$ we can rewrite Eq. (\ref{31}) as 
\begin{equation}
\left[ -i\hbar c\sqrt{1+\lambda r^{2}}\mathbf{\alpha }_{r}\left( \frac{%
\partial }{\partial r}+\frac{1}{r}-\frac{\beta }{\hbar r}\boldsymbol{K}%
\right) +\beta mc^{2}-E+V\right] \psi =0.
\end{equation}%
From this equation, we get two coupled differential equations:%
\begin{equation}
\left( \frac{E-mc^{2}}{\hbar c}+Z\mu \frac{\sqrt{1+\lambda r^{2}}}{r}\right)
f_{1}\left( r\right) =\sqrt{1+\lambda r^{2}}\left( \frac{\kappa -1}{r}-\frac{%
d}{dr}\right) f_{2}\left( r\right) ,  \label{40}
\end{equation}%
\begin{equation}
\left( \frac{E+mc^{2}}{\hbar c}+Z\mu \frac{\sqrt{1+\lambda r^{2}}}{r}\right)
f_{2}\left( r\right) =\sqrt{1+\lambda r^{2}}\left( \frac{\kappa +1}{r}+\frac{%
d}{dr}\right) f_{1}\left( r\right) .  \label{41}
\end{equation}
These two equations can be solved exactly by a diagonalization procedure 
\cite{36}, 
\begin{equation}
\left( 
\begin{array}{c}
g_{1} \\ 
g_{2}%
\end{array}%
\right) =\left( 
\begin{array}{cc}
1 & X \\ 
X & 1%
\end{array}%
\right) \left( 
\begin{array}{c}
f_{1} \\ 
f_{2}%
\end{array}%
\right) ,
\end{equation}%
where $X=\frac{\gamma -\kappa }{Z\mu }$ and $\gamma =\sqrt{\kappa
^{2}-\left( Z\mu \right) ^{2}}$. which transforms Eqs. (\ref{40}) and (\ref%
{41}) into the following equations, 
\begin{equation}
\left[ \sqrt{1+\lambda r^{2}}\frac{d}{dr}+\left( \gamma +1\right) \frac{%
\sqrt{1+\lambda r^{2}}}{r}-\frac{Z\mu }{\gamma }\frac{E}{\hbar c}\right]
g_{1}=\left[ \frac{mc^{2}}{\hbar c}+\frac{\kappa }{\gamma }\frac{E}{\hbar c}%
\right] g_{2},
\end{equation}%
\begin{equation}
\left[ \sqrt{1+\lambda r^{2}}\frac{d}{dr}+\left( 1-\gamma \right) \frac{%
\sqrt{1+\lambda r^{2}}}{r}+\frac{Z\mu }{\gamma }\frac{E}{\hbar c}\right]
g_{2}=\left[ \frac{mc^{2}}{\hbar c}-\frac{E}{\hbar c}\frac{\kappa }{\gamma }%
\right] g_{1}.
\end{equation}
This system gives the following differential equation for the component $%
g_{2}\left( r\right) =\frac{1}{\sqrt{r}}\Xi \left( r\right) ,$%
\begin{equation}
\left[ \left( 1+\lambda r^{2}\right) \left( \frac{d^{2}}{dr^{2}}+\frac{1}{r}%
\frac{d}{dr}\right) +\lambda r\frac{d}{dr}-\frac{\left( \gamma -\frac{1}{2}%
\right) ^{2}}{r^{2}}+\frac{2Z\mu E}{\hbar c}\frac{\sqrt{1+\lambda r^{2}}}{r}+%
\frac{E^{2}-m^{2}c^{4}}{\hbar ^{2}c^{2}}-\lambda \left( \gamma ^{2}-\frac{1}{%
4}\right) \right] \Xi \left( r\right) =0.
\end{equation}
Following the same procedure in the Klein-Gordon case, the solution is
obtained as by,%
\begin{equation}
\Xi =y^{a_{1}}\left( 1-y\right) ^{b_{1}}\mathbf{F}\left( A_{1};B_{1};\frac{1%
}{2}+2a,y\right) ,
\end{equation}%
with $\mathbf{F}$ being a hypergeometric function with the parameters%
\begin{equation}
A_{1}=a_{1}+b_{1}-\left( \gamma -\frac{1}{2}\right) ;\text{ \ }%
B_{1}=a_{1}+b_{1}+\left( \gamma -\frac{1}{2}\right) ,
\end{equation}
where%
\begin{equation}
a_{1}=\frac{1}{4}+\frac{1}{2}\sqrt{\gamma ^{2}-\frac{2\mu ZE}{\hbar c\sqrt{%
\lambda }}-\frac{E^{2}-m^{2}c^{4}}{\hbar ^{2}c^{2}\lambda }};\text{ \  \ }%
b_{1}=\frac{1}{4}+\frac{1}{2}\sqrt{\frac{2\mu ZE}{\hbar c\sqrt{\lambda }}%
+\gamma ^{2}-\frac{E^{2}-m^{2}c^{4}}{\hbar ^{2}c^{2}\lambda }}.
\end{equation}
If $A_{1}=-n$ or $B_{1}=-n,$ ($n=0,1,2,...$), the hypergeometric function
reduces to a polynomial in $y$ whose degree is $n,$ and then we have%
\begin{equation}
\gamma +\frac{1}{2}\sqrt{\gamma ^{2}+\frac{m^{2}c^{4}}{\hbar
^{2}c^{2}\lambda }-\frac{2\mu ZE}{\hbar c\sqrt{\lambda }}-\frac{E^{2}}{\hbar
^{2}c^{2}\lambda }}+\frac{1}{2}\sqrt{\gamma ^{2}+\frac{m^{2}c^{4}}{\hbar
^{2}c^{2}\lambda }+\frac{2\mu ZE}{\hbar c\sqrt{\lambda }}-\frac{E^{2}}{\hbar
^{2}c^{2}\lambda }}=-n.  \label{49}
\end{equation}%
We can obtain the energy eigenvalues by solving this equation for $E$,
namely, 
\begin{equation}
E_{Dirac}^{dS}=\frac{mc^{2}\sqrt{1-\frac{\hbar ^{2}\lambda }{m^{2}c^{2}}%
\left[ \left( N-j-\frac{1}{2}+\sqrt{\left( j+\frac{1}{2}\right) ^{2}-\left(
Z\mu \right) ^{2}}\right) ^{2}+\left( Z\mu \right) ^{2}-\left( j+\frac{1}{2}%
\right) ^{2}\right] }}{\sqrt{1+\frac{\mu ^{2}Z^{2}}{\left( N-j-\frac{1}{2}+%
\sqrt{\left( j+\frac{1}{2}\right) ^{2}-\left( Z\mu \right) ^{2}}\right) ^{2}}%
}},  \label{50}
\end{equation}
where $N=n+j+\frac{1}{2}$ is the principal quantum number. The equation (\ref%
{50}) gives the energy levels of hydrogen-like atoms in de Sitter space. The
energy levels depend on the principal quantum number $N$, $j$, $Z$ and the
deformation parameter $\lambda $. It should be emphasized that the energy
depends on the quantum number $j$, which is associated with both orbital
angular momentum and spin. The other thing to notice about Eq. (\ref{50}) is
that for states with $N=1$ and $j=1/2,$ the energy spectrum is given by, 
\begin{equation}
E_{Dirac}^{dS}=mc^{2}\sqrt{1-\left( Z\mu \right) ^{2}}.
\end{equation}%
In this special limit, it can be easily seen that the effects of the
deformation have been totally disappeared and, if $\left( Z\mu \right) =1,$
the energy becomes zero, i.e. $E_{N}^{ds}=0$.

In addition, if $Z\mu >j+\frac{1}{2}$ the energy levels in (\ref{50}) become
complex which means that there exists no regular polynomial solutions for $%
ns_{1/2}$ or $np_{1/2}$ states when the charge is greater than $Z>137$.

Expanding Eq.(\ref{50}) to the first order in $\lambda $, we obtain%
\begin{equation}
E_{Dirac}^{dS}=\varepsilon _{Dirac}-\frac{\hbar ^{2}\lambda \varepsilon
_{Dirac}}{2m^{2}c^{2}}\left[ \left( N-j-\frac{1}{2}+\sqrt{\left( j+\frac{1}{2%
}\right) ^{2}-\left( Z\mu \right) ^{2}}\right) ^{2}+\left( Z\mu \right)
^{2}-\left( j+\frac{1}{2}\right) ^{2}\right] ,  \label{52}
\end{equation}
where%
\begin{equation}
\varepsilon _{Dirac}=mc^{2}\left[ 1+\frac{\mu ^{2}Z^{2}}{\left( N-j-\frac{1}{%
2}+\sqrt{\left( j+\frac{1}{2}\right) ^{2}-\left( Z\mu \right) ^{2}}\right)
^{2}}\right] ^{\frac{-1}{2}}.
\end{equation}
Cleary, the first term is identical with the Dirac energy levels in the
ordinary case. The second term represents the correction due to the presence
of the extended uncertainty principle. Moreover, the series expansion of (%
\ref{52}) in powers of $\left( Z\mu \right) $ reads%
\begin{equation}
W_{Dirac}^{ds}=E_{N}^{dS}-mc^{2}=\epsilon _{N;j}+\Delta \mathcal{E}_{N;j},
\label{54}
\end{equation}%
where 
\begin{equation}
\epsilon _{N;j}=-\frac{Z^{2}e^{4}m}{2\hbar ^{2}N^{2}}-mc^{2}\frac{\mu
^{4}Z^{4}}{2N^{4}}\left( \frac{N}{\left( j+\frac{1}{2}\right) }-\frac{3}{4}%
\right).
\end{equation}
Here, the first term is the energy spectrum of the non-relativistic hydrogen
atom, the second term contains all of the details of the fine structure, and%
\begin{equation}
\Delta \mathcal{E}_{N;j}=-\frac{\hbar ^{2}\lambda }{2m}\left[ 1-\frac{%
Z^{2}\mu ^{2}}{2N^{2}}-\frac{\mu ^{4}Z^{4}}{2N^{4}}\left( \frac{N}{j+\frac{1%
}{2}}-\frac{3}{4}\right) \right] \left[ N^{2}-\left( j+\frac{1}{2}\right)
^{2}-\left( Z\mu \right) ^{2}\left( 1+\frac{\left( Z\mu \right) ^{2}}{%
4\left( j+\frac{1}{2}\right) ^{2}}\right) \left( \frac{N}{j+\frac{1}{2}}%
-1\right) \right].
\end{equation}
The formula (\ref{54}) shows the effect of the EUP on the non-relativistic
energy levels for the one-electron atom. We observe that the EUP correction
carries new terms associated with the relativistic correction, which do not
exist in the undeformed case.

Consequently, we can calculate the electron energies of hydrogen-like atoms
in de Sitter space using our results with the numerical values of $%
mc^{2}=511004.1\unit{eV}$ and $\sqrt{\lambda }=0.252\times 10^{6}\unit{m}%
^{-1}$ \cite{28,37}$.$ The results are given in Table 1.

\begin{center}
\bigskip 
\begin{tabular}{|l|l|l|l|l|l|}
\hline
$N$ & $\ell $ & $j$ & Label & $\epsilon _{N;j}$ \ ($\unit{eV}$) & $\left
\vert \Delta \mathcal{E}_{N;j}\right \vert $ \ ($\unit{eV}$) \\ \hline
1 & 0 & 1/2 & $1s_{1/2}$ & $-13.605$ & $0$ \\ \hline
2 & 0 & 1/2 & $2s_{1/2}$ & $-3.40132$ & $3.\, \allowbreak 415\,0\times
10^{-8} $ \\ \hline
2 & 1 & 1/2 & $2p_{1/2}$ & $-3.40132$ & $3.\, \allowbreak 415\,0\times
10^{-8} $ \\ \hline
2 & 1 & 3/2 & $2p_{13/2}$ & $-3.40127$ & $0$ \\ \hline
3 & 0 & 1/2 & $3s_{1/2}$ & $-1.51169$ & $1.\, \allowbreak 940\,5\times
10^{-8} $ \\ \hline
3 & 1 & 1/2 & $3p_{1/2}$ & $-1.51169$ & $1.\, \allowbreak 940\,5\times
10^{-8} $ \\ \hline
3 & 1 & 3/2 & $3p_{3/2}$ & $-1.51168$ & $1.\, \allowbreak 212\,8\times
10^{-8} $ \\ \hline
3 & 2 & 3/2 & $3d_{3/2}$ & $-1.51168$ & $1.\, \allowbreak 212\,8\times
10^{-8} $ \\ \hline
3 & 2 & 5/2 & $3d_{5/2}$ & $-1.51167$ & $0$ \\ \hline
\end{tabular}

\begin{tabular}{l}
Table 1: Energy levels of hydrogen-like atoms.%
\end{tabular}
\end{center}

The same calculation can be performed for the case of anti-de Sitter space.
According to the correspondence $\lambda \rightarrow -\lambda $, the energy
levels of a spin-1/2 particle in a Coulomb potential in anti-de Sitter space
can be written as 
\begin{equation}
E_{Dirac}^{AdS}=\frac{mc^{2}\sqrt{1+\frac{\hbar ^{2}\lambda }{m^{2}c^{2}}%
\left[ \left( N-j-\frac{1}{2}+\sqrt{\left( j+\frac{1}{2}\right) ^{2}-\left(
Z\mu \right) ^{2}}\right) ^{2}+\left( Z\mu \right) ^{2}-\left( j+\frac{1}{2}%
\right) ^{2}\right] }}{\sqrt{1+\frac{\mu ^{2}Z^{2}}{\left( N-j-\frac{1}{2}+%
\sqrt{\left( j+\frac{1}{2}\right) ^{2}-\left( Z\mu \right) ^{2}}\right) ^{2}}%
}}.
\end{equation}
We can plot our results as in the Klein-Gordon case, following the same
numerical procedure. We observe the same behavior in the energy curves.

\begin{figure}[tbph]
\centering
\includegraphics[scale=0.5]{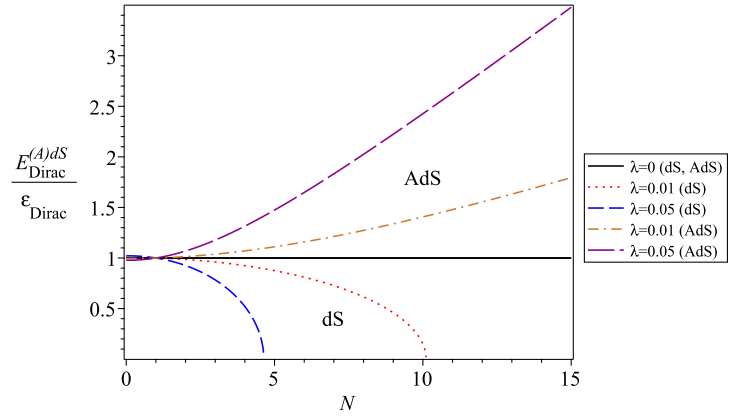}
\caption{$\frac{E_{Dirac}^{(A)dS}}{\protect \varepsilon_{Dirac}}$ vs. $N$}
\label{fig:fig4}
\end{figure}

In Fig. (\ref{fig:fig4}) we plot the ratio $\frac{E_{Dirac}^{(A)dS}}{%
\varepsilon_{Dirac}}$ with respect to $N$ for some values of $\lambda$. We
see that the curves for the AdS case grow with $N$, while they decrease with 
$N$ in the dS case. The undeformed energy value corresponding to $\lambda=0$
is shown with a solid line where $E_{Dirac}^{(A)dS}=\varepsilon_{Dirac}$.

\begin{figure}[tbph]
\centering
\includegraphics[scale=0.5]{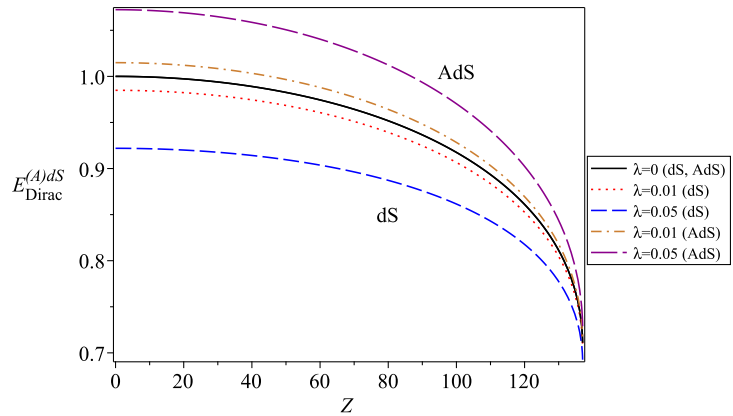}
\caption{$E_{Dirac}^{(A)dS}$ vs. $Z$ for $N=2$, $j=1/2$}
\label{fig:fig5}
\end{figure}

\begin{figure}[tbph]
\centering
\includegraphics[scale=0.5]{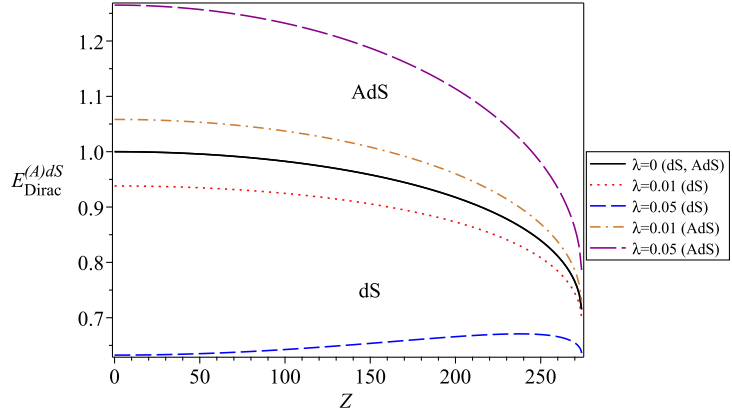}
\caption{$E_{Dirac}^{(A)dS}$ vs. $Z$ for $N=4$, $j=3/2$}
\label{fig:fig6}
\end{figure}

The Figures (\ref{fig:fig5}) and (\ref{fig:fig6}) show the behavior of the
energy eigenvalues $E_{Dirac}^{(A)dS}$ with respect to $Z$. In Fig. (\ref%
{fig:fig5}) we take $N=2$ and $j=1/2$ and using $Z \mu>j+\frac{1}{2}$, we
see that the bound states are limited by $Z<137$. The results are similar in
Fig. (\ref{fig:fig6}) where $N=4$, $j=3/2$ and $Z<274$. 

\section{Conclusion}

We studied the Klein-Gordon and Dirac equations for the relativistic Coulomb
potential in the framework of the extended uncertainty principle for de
Sitter and anti-de Sitter spaces. We treated the problem analytically and
obtained the energy spectra and the eigenfunctions exactly in both cases.
Eventually, we obtained the relativistic bound-state energy spectra for
hydrogenic atoms and the corresponding wave functions for the spin $0$ and $%
1/2$ cases.

We solved the Klein-Gordon and Dirac equations in terms of the
hypergeometric functions and utilizing the conditions that truncate the
series to yield a polynomial solution, we obtained the energy eigenvalues.
Analyzing the energy spectra, we found the limits $\ell +\frac{1}{2}%
\geqslant Z\mu $ and $Z\mu <j+\frac{1}{2}$ which are necessary for
the existence of physical energy eigenvalues in the Klein-Gordon and Dirac
cases, respectively.

Furthermore, we used our analytical results to create a numerical table for
the energy levels of hydrogen-like atoms and plot some graphics to present
the effects of the deformation on the energy spectra and the accumulation of
the curves to a certain $Z$-value in a visual way for de Sitter and anti-de
Sitter spaces.


\begin{thebibliography}{99}
\bibitem{1} W. Heisenberg, Z. Phys. 43, 172 (1927).

\bibitem{2} J. Lukierski, H. Ruegg, A. Nowicki, and V. N. Tolstoy, Phys.
Lett. B 264, 331 (1991).

\bibitem{3} J. Lukierski, A. Nowicki, and H. Ruegg, Phys. Lett. B 293, 344
(1992).

\bibitem{4} J. Magueijo and L. Smolin, Phys. Rev. Lett. 88, 190403 (2002).

\bibitem{5} G. Amelino-Camelia, Int. J. Mod. Phys. D 11, 35 (2002).

\bibitem{6} G. Amelino-Camelia, Phys. Lett. B 510, 255 (2001).

\bibitem{7} D. J. Gross and P. F. Mende, Nucl. Phys. B 303, 407 (1988).

\bibitem{8} O. Aharony, S. S. Gubser, J. Maldacena, H. Ooguri and Y. Oz,
Physics Reports 323, 183 (2000).

\bibitem{9} M. Faizal ,Int. J. Mod. Phys. A 29, 1450106 (2014).

\bibitem{10} M. Faizal, B. Majumder, Ann. Phys., NY 357 49 (2015).

\bibitem{11} L. J. Garay, Int. J. Mod. Phys. A 10, 145 (1995).

\bibitem{12} F. Scardigli, Phys. Lett. B 452, 39 (1999).

\bibitem{13} A. Bina, S. Jalalzadeh and A. Moslehi, Phys. Rev. D 81, 023528
(2010).

\bibitem{14} A. Kempf, G. Mangano, and R. B. Mann, Phys. Rev. D 52, 1108
(1995).

\bibitem{15} A. F. Ali, S. Das, and E. C. Vagenas, Phys. Rev. 84, 1 (2011).

\bibitem{16} S. Das and R. B. Mann, Phys. Lett. B 704, 596 (2011).

\bibitem{17} S. Mignemi, Mod. Phys. Lett. A 25, 1697 (2010).

\bibitem{18} R.N. Costa Pilho, J.P.M. Braga, J.H.S. Lira, J.S. Anrade, Phys.
Lett. B 755, 367 (2016).

\bibitem{19} J.R. Mureika, Phys. Lett. B 789, 88 (2019).

\bibitem{20} T. Schurmanna, Eur. Phys. J. C 80, 141 (2020).

\bibitem{21} M. P. Dabrowski, F. Wagner, arXiv:2006.02188.

\bibitem{22} M.P. Dabrowski, F. Wagner, Eur. Phys. J. C 79, 716 (2019).

\bibitem{23} B. Hamil and M. Merad, Eur. Phys. J. Plus 133, 174 (2018).

\bibitem{24} N. Messai, B. Hamily and A. Hafdallah, Mod. Phys. Lett. A 34,
1850202 (2018) .

\bibitem{25} B. Hamil, Indian J. Phys. 93(10), 1319 (2019).

\bibitem{26} B. Hamil and M. Merad, Int. J. Mod. Phys. A 33, 1850177 (2018).

\bibitem{27} B. Hami, M. Merad and T. Birkandan, Int. J. Mod. Phys. A 35,
2050014 (2020).

\bibitem{28} B. Hamil and M. Merad, Few-Body Syst. 60, 36 (2019).

\bibitem{29} B. Hamil, M. Merad and T. Birkandan, Eur. Phys. J. Plus 134,
278 (2019).

\bibitem{30} B. Hamil,M. Merad and T. Birkandan, Phys. Scr. 95, 075309
(2020).

\bibitem{31} S. Ghosh, S. Mignemi, Int. J. Theor. Phys. 50, 1803 (2011).

\bibitem{32} W. S. Chung, H. Hassanabadi and N. Farahani, Mod. Phys. Lett. A
34, 1950204 (2019).

\bibitem{33} W.S. Chung, H. Hassanabadi, Mod. Phys. Lett. A 32, 1750138
(2017).

\bibitem{34} O. Yesiltas, Eur. Phys. J. Plus 134, 331 (2019).

\bibitem{35} M. Falek, N. Belghar and M. Moumni, Eur. Phys. J. Plus 135, 335
(2020).

\bibitem{36} R. A. Swainson and G. W. F. Drake, J. Phys. A: Math. Gen. 24,
79 (1991).

\bibitem{37} M. Zarei and B. Mirza, Phys. Rev. D 79, 125007 (2009).
\end{thebibliography}
\end{document}